\PassOptionsToPackage{numbers, sort&compress}{natbib}
\documentclass{article}

\usepackage[final, main]{neurips_2026}

\usepackage[utf8]{inputenc} 
\usepackage[T1]{fontenc}    
\usepackage[dvipsnames]{xcolor}         
\usepackage{hyperref}
\hypersetup{
    colorlinks=true,
    urlcolor=blue,
    citecolor=blue,
    linkcolor=blue,
}
\usepackage{url}            
\usepackage{booktabs}       
\usepackage{float}
\usepackage{siunitx}
\usepackage{multirow}
\usepackage{amsfonts}       
\usepackage{nicefrac}       
\usepackage{microtype}      
\usepackage{amsmath}
\usepackage{graphicx}
\usepackage[labelfont=bf]{caption}

\title{Sampling the Schwinger Model with Gauge-Equivariant Diffusion}

\author{
  Octavio Vega
  \qquad
  Aida X. El-Khadra \\
  Department of Physics \\
  University of Illinois Urbana-Champaign \\
  Urbana, IL 61801 \\
  \href{mailto:octavio5@illinois.edu}{\texttt{octavio5@illinois.edu}}
}

\begin{document}

\maketitle

\begin{abstract}
  We present a first study of a diffusion-based approach to accelerated sampling of the $N_f = 2$ lattice Schwinger model. Our work is inspired by recent and growing successes in developing such generative models for ensemble generation in LFT to overcome the well-known critical slowing down problem~\cite{wang:2023, zhu:2025, vega:2025, kanwar:2025, aarts:2026, alharazin:2026, komijani:2026}. We train a U(1)-equivariant score-based generative model to sample gauge link configurations from the marginal Schwinger model. By computing model likelihoods, we obtain unbiased estimates for observables that closely match those produced by MCMC simulations. We also demonstrate improvement over HMC as measured qualitatively by a reduction in topological freezing near critical parameters.
\end{abstract}

\section{Introduction}
Across many domains in physics, quantum field theory (QFT) is a useful tool for studying interacting systems. Lattice field theory (LFT) offers a non-perturbative avenue for simulating these systems by discretizing continuum theories onto a Euclidean spacetime grid. LFT is based on the path integral formalism of QFT, where physical observables are computed as weighted expectations:
\begin{equation}\label{eq:path_integral_obs}
    \langle \mathcal{O}\rangle = \frac{1}{\mathcal{Z}}\int \left[\mathcal{D}U\right] \, \mathcal{O}(U) e^{-S[U]} \approx \frac{1}{N} \sum_{i=1}^N \mathcal{O}(U_i).
\end{equation}
The integration over all possible configurations of the field $U$ is specified by the functional measure $\left[\mathcal{D}U\right]$. In LFT, numerical simulations are used to approximate the expectation values of Eq.~\eqref{eq:path_integral_obs} via empirical means as in the right-most expression of Eq.~\eqref{eq:path_integral_obs}, which amounts to sampling large ensembles of field configurations $\{U_i\}_{i=1}^{N}$ from the physical distribution $p(U) = \nicefrac{e^{-S[U]}}{\mathcal{Z}}$. 

As LFT simulations approach the continuum limit of the underlying models, Markov chain Monte Carlo (MCMC) sampling algorithms become severely hindered, a phenomenon known as \emph{critical slowing down}~\cite{Wolff1990}. Autocorrelation times between samples grow exponentially large due to low acceptance rates between new proposals. At the same time, for topologically non-trivial theories, samples can become ``frozen" in particular modes and thereby fail to explore the various topological sectors of configuration space~\cite{woit:1983, alles:1996}.

In this paper we study the Schwinger model~\cite{schwinger1962}, a fermionic QFT representing quantum electrodynamics in two dimensions. It is an effective toy model for quantum chromodynamics (QCD), as it exhibits confinement, a chiral anomaly, and non-trivial topology. 
Previous work showed that normalizing flows~\cite{Rezende2015, Dinh2017, Papamakarios2021} can yield faster approaches to sampling compared to Hybrid Monte Carlo (HMC)~\cite{duane:1987} in the lattice Schwinger model~\cite{AlbergoSchwingerFlows2022, FinkenrathSchwingerFlows2022}. In this work, we present another method rooted in diffusion models. To our knowledge, this is the first application of diffusion-based sampling to a lattice gauge theory with fermions.

\section{Diffusion Models}
Diffusion models~\cite{Sohl-Dickstein:2015, Ho:2020, Song:2021} are a type of generative model formulated through stochastic differential equations (SDE). The \emph{forward process}, relevant for training, is responsible for corrupting initial configurations by repeatedly injecting increasing levels of random noise over diffusion time. The \emph{reverse process}, used to generate new samples, employs a trained score network to interpolate from noise to data by integrating an SDE along reverse diffusion time.

The forward process is captured by the variance-expanding SDE, written in Itô form as
\begin{equation}\label{eq:forward_SDE}
    d\phi_t = g_tdW_t,
\end{equation}
and describes the progressive corruption of initial configurations $\phi_0$ towards noise. The diffusion coefficient $g_t: [0, 1] \to \mathbb{R}^+$ is a thermal diffusivity controlling the level of noise added to samples over time. When adapted to the Lie group U(1), the analytical solution to Eq.~\eqref{eq:forward_SDE} yields
\begin{equation}
    \phi_0 \mapsto \phi_t = \operatorname{wrap}\left(\phi_0 + \sigma_t \eta\right), \quad  \sigma_t = \sqrt{\int_0^t g_s^2 ds}
\end{equation}
where $\eta \sim \mathcal{N}(0, \mathbb{I})$ and the $\operatorname{wrap}$ operation folds into the unit circle modulo $2\pi$.

The SDE which reverses the forward process in Eq.~\eqref{eq:forward_SDE} is itself a diffusion process:
\begin{equation}\label{eq:reverse_SDE}
    d\phi_t = -\frac{g_t^2 + \tilde{g}_t^2}{2}\nabla \log p_t(\phi_t) + \tilde{g}_t dW_t.
\end{equation}
The gradient field $\nabla \log p_t$ is known as the \emph{score function}, and can be estimated by a time-dependent neural network $\hat{\boldsymbol{s}}_t$ trained through score matching~\cite{Song:2019}. The parameters of $\hat{\boldsymbol{s}}_t$ are optimized by minimizing the conditional score matching objective:
\begin{equation}
    \mathcal{L}\left[\hat{\boldsymbol{s}}\right] = \int_0^1 \mathbb{E}_{\phi_0 \sim p_0} \mathbb{E}_{\phi_t \sim p_{t|0}}\left[\lambda_t\left\|\hat{\boldsymbol{s}}_t(\phi_t) - \nabla \log p_{t|0}(\phi_t | \phi_0)\right\|^2\right]dt
\end{equation}
where $\lambda_t$ is a positive weight.
Then, the reverse process \eqref{eq:reverse_SDE} can be simulated using a numerical ODE solver to obtain new samples. In order to calculate the likelihoods of samples under the model, we set $\tilde{g}_t \equiv 0$ and integrate the deterministic ODE in reverse as in continuous normalizing flows~\cite{Chen:2018}.

\section{The Schwinger Model on the Lattice}
\subsection{Definitions}
The action for the lattice Schwinger model is given as
\begin{equation}\label{eq:schwinger_joint_action}
    S[U, \psi, \bar{\psi}] = -\beta \sum_x \mathrm{Re} \, P(x) + \sum_{x, y} \bar{\psi}_\alpha(x) D[U](x, y)^{\alpha\beta}\psi_\beta(y),
\end{equation}
where $P(x) = U_0(x)U_1(x + \hat{0}) U_0^*(x + \hat{1}) U_1^*(x)$ is the U(1) plaquette. Distinct lattice sites are denoted by $x, y$ and $\alpha, \beta \in \{1, 2\}$ are spin indices that are summed. The differential operator $D[U]$ is the Wilson Dirac operator, given by
{\small 
\begin{align}\label{eq:dirac_op}
    D[U](x, y)^{\alpha\beta} = \delta_{x, y}\delta^{\alpha\beta} - \kappa\sum_\mu \bigl\{U_\mu^*(y) \delta_{x, y + \hat{\mu}} \left[\mathbb{I} + \sigma_\mu\right]^{\alpha\beta}
     + \; U_\mu(y - \hat{\mu}) \delta_{x, y - \hat{\mu}} \left[\mathbb{I} - \sigma_\mu\right]^{\alpha\beta}\bigr\}.
\end{align}
}The matrices $\sigma_\mu$ for $\mu = 1, 2$ are the Hermitian $2\times2$ Pauli matrices and $\mathbb{I}$ is the identity matrix in spin space. The strength of the coupling between the bosonic variables $U$ and the fermions $\psi$ is controlled by the \emph{Hopping parameter} $\kappa$, related to the bare fermion mass $m_0$ via $\kappa = \tfrac{1}{4 + 2m_0}$. 

The full action \eqref{eq:schwinger_joint_action} defines a joint probability $p\left(U, \psi, \bar{\psi}\right)$ over the gauge links and fermions which can be marginalized by integrating out the fermion fields~\cite{Gattringer1997, GattringerLang2010}, giving a marginal density
\begin{equation}
    p(U) \propto \det\left(D[U]^\dagger D[U]\right) e^{-S_g[U]}, \quad S_g[U] = -\beta \sum_x \operatorname{Re}P(x).
\end{equation}
In turn, the marginal density defines an effective action $S_{\rm eff}[U] = -\log p(U)$ through
\begin{equation}\label{eq:exact_det}
    S_{\rm eff}[U] = -\beta \sum_x \mathrm{Re} \, P(x) - \log \det \left(D[U]^\dagger D[U]\right).
\end{equation}
Since the complex-valued gauge links are parameterized as $U = \exp(iA)$, we work directly with their phases $A \in [0, 2\pi)$, which are dimensionless in lattice units. Our task is then to sample gauge link angles $A$ from the marginal density defined by the exact-determinant action in Eq.~\eqref{eq:exact_det}.

\subsection{Computational Setup}
For our simulations, we investigate two square lattice sizes: $L^2 = 8 \times 8$ and $L^2 = 16 \times 16$. We consider $N_f = 2$ flavors of degenerate fermions with the action parameters $\beta = 2.0$ and $\kappa = 0.276$. Temporal anti-periodic boundary conditions are imposed inside the delta functions of Eq.~\eqref{eq:dirac_op}. Our training datasets consist of 8192 gauge link configurations generated using HMC with 200 trajectories each consisting of 10 leapfrog steps at step size 0.1.

The prior distribution is taken to be uniform with respect to the Haar measure, which for U(1) is $\nicefrac{1}{2\pi}$ on $[0, 2\pi)$ and 0 elsewhere. The diffusion schedule is chosen to be linear in time: $g_t =3t$, yielding a marginal width $\sigma_t = \sqrt{3t^3}$.

Models are trained for 1000 epochs in batches of 512 with a learning rate of 0.004 using the Adam optimizer~\cite{kingma:2015}. The score network uses a U-Net parameterization~\cite{ronneberger:2015} incorporating time encoding via Gaussian Fourier embeddings~\cite{tancik:2020}. For inference, we integrate the reverse ODE using an Euler integrator with 500 steps and step size 0.002. Likelihoods are obtained using the Skilling-Hutchinson trace estimator~\cite{skilling:1989, hutchinson:1989} with 10 Rademacher random vectors.

The high degree of non-locality introduced by the fermion determinant demands architectural features with greater expressive potential. To that end, we have thus far experimented with dilated convolutions~\cite{Yu:2015} as well as axial attention layers~\cite{Ho:2019} to learn long-range correlations and expand the receptive field of our network. Moreover, in Ref.~\cite{vega:2025} we showed how to construct a U(1) gauge-equivariant score network by feeding it gauge-invariant plaquettes. Here, we use rectangular $2 \times 1$ Wilson loops as additional gauge-invariant input channels to provide greater coverage over the lattice, as was done in Ref.~\cite{AlbergoSchwingerFlows2022}.

\section{Results}
\subsection{Sampling Dynamics}
\begin{figure}[h]
    \centering
    \includegraphics[width=0.65\linewidth]{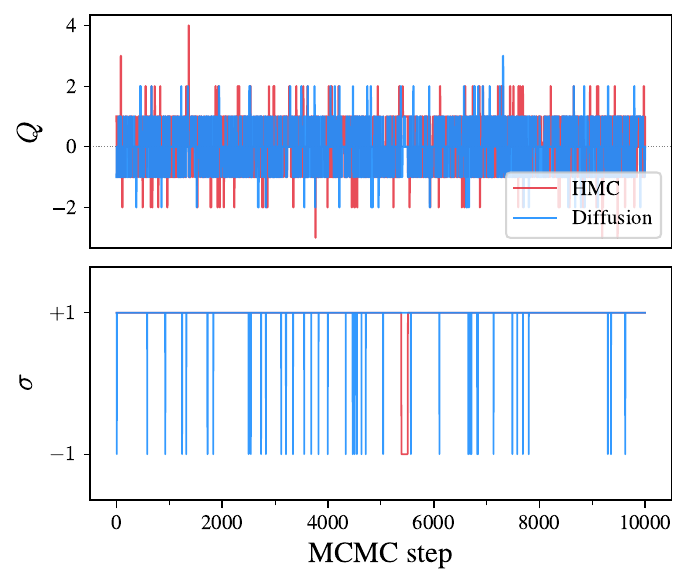}
    \caption{Evolution of the topological charge $Q$ and Dirac sign $\sigma$ during sampling on an $8 \times 8$ lattice.}
    \label{fig:sampling_dynamics}
\end{figure}

The integer-valued topological charge is used to label topological sectors and is defined as
\begin{equation}\label{eq:topo_charge}
    Q  = \frac{1}{2\pi} \sum_x \arg P(x) \in \mathbb{Z}.
\end{equation}
Another observable was suggested in Ref.~\cite{Gattringer1997} to identify tunneling events between even and odd topological sectors, given by
\begin{equation}
    \sigma = \mathrm{sign} \left( \mathrm{Re} \, \det D \right).
\end{equation}
For comparison, we resample~\cite{doucet:2001} the diffusion-generated configurations using 10,000 Metropolis accept/reject steps~\cite{Metropolis:1953} and generate an HMC dataset with the same number of steps. We track both $Q$ and $\sigma$ during MCMC time for both HMC and diffusion, as depicted in Figure~\ref{fig:sampling_dynamics}.
Though the evolution of $Q$ is comparable, we observe significantly more even-odd transitions with the diffusion model when compared to HMC, suggesting that diffusion-based sampling can enable much wider and more efficient phase space exploration.

\subsection{Observables}
We validate the diffusion-generated configurations by computing physical observables on them and comparing them to those computed over a reference dataset generated with HMC.

Two observables carry over from pure gauge theory, which are the average value of the volume-normalized plaquette $\langle P \rangle =  L^{-2} \langle \operatorname{Re} \sum_x P(x) \rangle$ and the topological susceptibility $\chi_Q = \langle Q^2\rangle$. We also investigate a fermionic observable known as the chiral condensate $\langle \bar{\psi}\psi\rangle$, given by the volume-averaged trace of $D^{-1}[U]$.

In Table~\ref{tab:results}, we show the effective sample size (ESS)~\cite{Liu:1996}, MCMC acceptance rate (AR), and the unnormalized Kullback-Leibler divergence (KL)\footnote{The reported KL is $\mathbb{E}_{U \sim q} \left[S[U] + \log q(U)\right]$, an estimate of $D_{\rm KL}(q \| p) - \log\mathcal{Z}$, which allows for negative values.} along with preliminary measurements of the three observables across the two lattice sizes for both HMC and diffusion.

\begin{table}[H]
\centering
    \setlength{\tabcolsep}{5pt}
    \begin{tabular}{c l ccc S[table-format=2.2] S[table-format=1.4(2)] S[table-format=1.4(2)] S[table-format=1.3(2)]}
        \toprule
        & & \multicolumn{4}{c}{\textbf{Metrics}} & \multicolumn{3}{c}{\textbf{Observables}} \\
        \cmidrule(lr){3-6} \cmidrule(l){7-9}
        $L$ & \textbf{Method} & {ESS} $\uparrow$ & AR $\uparrow$ & KL $\downarrow$ & {$\tau_{\rm int}(Q)$ $\downarrow$} & {$\langle P \rangle$} & {$\chi_Q$} & {$\langle\bar{\psi}\psi\rangle$} \\
        \midrule
        \multirow{2}{*}{8}  & HMC       & {--} & {--} & {--}      & {1.22}  & 0.7597(14) & 0.0038(03) & 1.502(04) \\
                            & Diffusion & 0.29 & 35\% & $-295.0$  & {2.92}  & 0.7607(32) & 0.0037(05) & 1.504(14) \\
        \addlinespace
        \multirow{2}{*}{16} & HMC       & {--} & {--} & {--}      & 12.95 & 0.7585(08) & 0.0040(02) & 1.507(04) \\
                            & Diffusion & 0.08 & 20\% & $-1180.2$ & 4.91  & 0.7585(46) & 0.0035(14) & 1.501(25) \\
        \bottomrule
    \end{tabular}
    \vspace{2.0mm}
\caption{Quality metrics and observables computed on 1024 samples generated from HMC and diffusion. Diffusion model observables are reweighted.}
\label{tab:results}
\end{table}

\section{Outlook}
We present a first exploratory study of diffusion-based sampling in fermionic theories, demonstrating a new step forward for deep generative models in lattice gauge theory. 
Further experimentation across model architectures is needed to identify a scalable approach to sampling with fermions. Future work will investigate adaptations of diffusion-based sampling to different schemes for handling fermions~\cite{albergo:2021}, including with pseudofermions both in the Schwinger model and in ${\rm SU}(N)$ lattice gauge theory~\cite{abbott:2022}.

\begin{ack}
We thank Gurtej Kanwar, Javad Komikani, and Marina Marinkovic for helpful discussions.
AXK and OV acknowledge support from the U.S. Department of Energy, Office of Science under grant Contract Number DE-SC0015655. OV is grateful for support from the U.S. Department of Energy, Office of Science, “High Energy Physics Computing Traineeship for Lattice Gauge Theory” Award No. DESC0024053, from a UIUC Graduate College Fellowship, and from a Sloan Scholarship through the Alfred P. Sloan Foundation.
The numerical simulations are based on the \textbf{PyTorch}~\cite{paszke:2019} and \textbf{NumPy}~\cite{harris:2020} libraries, and figures are created with \textbf{Matplotlib}~\cite{hunter:2007}.
The results presented in this paper are obtained from allocations on the DeltaAI advanced computing and data resource, which is supported by the National Science Foundation (award OAC2320345) and the State of Illinois.
\end{ack}


\end{document}